\documentclass{article}

\usepackage{PRIMEarxiv}

\usepackage[utf8]{inputenc} % allow utf-8 input
\usepackage[T1]{fontenc}    % use 8-bit T1 fonts
\usepackage{hyperref}       % hyperlinks
\usepackage{url}            % simple URL typesetting
\usepackage{booktabs}       % professional-quality tables
\usepackage{amsfonts}       % blackboard math symbols
\usepackage{nicefrac}       % compact symbols for 1/2, etc.
\usepackage{microtype}      % microtypography
\usepackage{lipsum}
\usepackage{fancyhdr}       % header
\usepackage{graphicx}       % graphics
\graphicspath{{media/}}     % organize your images and other figures under media/ folder
\usepackage{xcolor}
\title{DCZNMaker: A Web-based Application for Multi-Attribute Utilities Analysis
\thanks{\textit{\underline{Citation}}: 
\textbf{Kline, DCZNMaker: A Web-based Application for Multi-Attribute Utilities Analysis, ArXiv, 2024}} 
}

\author{
  Adrienne Kline\\
  Center for Artificial Intelligence, BCVI, Northwestern Medicine\\
  Division of Cardiac Surgery, Northwestern University\\
  Xtasis Inc.\\
  \texttt{\ adrienne.kline@northwestern.edu} \\
}

\begin{document}
\maketitle

\begin{abstract}
DCZNMaker is a web-based application designed to streamline decision-making processes using
Multi-attribute Utility Analysis (MAUA). Built with simplicity and efficiency in mind, DCZNMaker
empowers users to make informed decisions among alternatives (options) by making explicit the factors (attributes) to be taken into consideration, as well as the importance (weights) and utility (location) of each attribute. The app offers a user-friendly interface, allowing individuals to input the various attributes and their associated weights and locations effortlessly. Leveraging advanced algorithms, DCZNMaker computes and presents comprehensive analyses, aiding users in understanding the relative importance of each attribute and guiding them towards optimal decisions. Several use cases are demonstrated. Whether for personal, professional, or academic use, DCZNMaker is a versatile tool adaptable to diverse decision-making scenarios. With its intuitive design and robust functionality, DCZNMaker revolutionizes decision-making processes, empowering individuals or groups of users to make well-informed choices with confidence and clarity. 
\end{abstract}

% keywords can be removed
\keywords{decision-making \and multiattribute utility analysis}

\section{Introduction}
\begin{figure}
    \centering
    \includegraphics[width=0.8\textwidth]{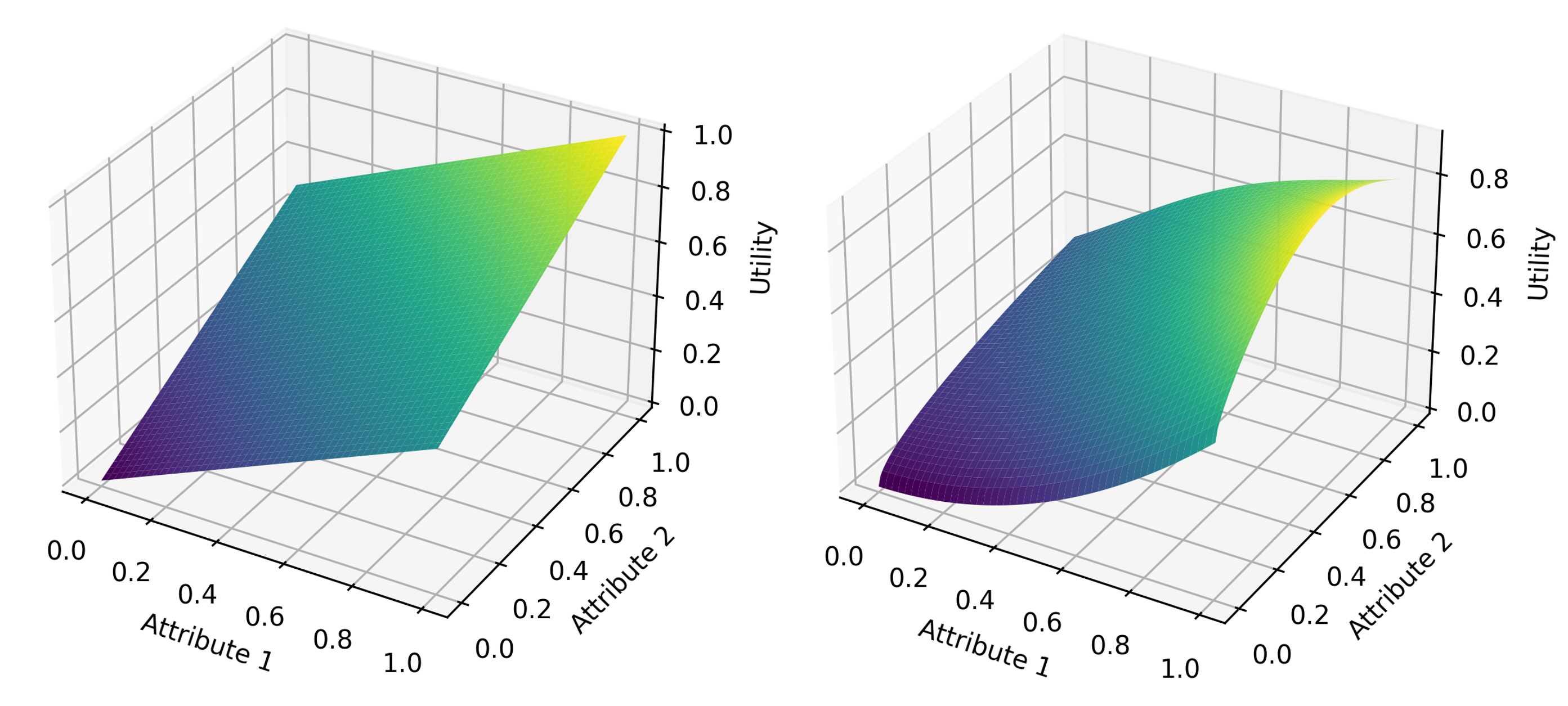}
    \caption{Visualization of decision-making surface for a two attribute decision paradigms. The graph on the left demonstrates a linear decision surfaces using two attributes, and the one on the right a non-linear decision surface.}
    \label{fig:decision surface}
\end{figure}

Multi-attribute utility analysis (MAUA) is a decision-making tool used in fields such as economics, business, and engineering to evaluate and compare multiple alternatives based on several criteria or attributes \cite{keeney1993decisions}. Multi-attribute utility analysis (MAUA) is a decision-making paradigm that employs an analytic hierarchy process \cite{saaty2008decision}\cite{reilly2001making}. The first step is to clearly articulate the objective of the decision (e.g., purchase one vehicle over others; select one process over others). The second step is to identify all possible options available for the decision at hand. These options could be vehicles, projects, investments, or any other decision situation where there are a finite number of choices. The next step is to determine the relevant attributes that will be used to evaluate the alternatives. These attributes could include cost, quality, reliability, durability, customer satisfaction and/or any other factors on which the decision will have an important consequence. Once the attributes are identified, in the fourth step weights are assigned to each attribute, reflecting the importance of each attribute in the decision-making process. For example, if cost is more important than quality, it would be assigned a higher weight. After the importance scores for each attribute are collected, they are converted into weights by dividing each attribute’s score by the total sum of all scores for each respondent \cite{stillwell1987comparing}. This results in individual weights for each attribute that sum to 1, as per Multi-Attribute Utility Theory \cite{stillwell1987comparing}. For example, if a respondent’s importance ratings for eight different attributes are 20, 30, 40, 50, 30, 60, 70, and 20, the total sum is 320. The weight for the first attribute, then, is 20/320 = 0.06, with other weights calculated similarly. The fifth step is to link each attribute to each alternative through locator values (utility). These values answer the question: “To what degree would this alternative on this attribute contribute to the overall objective of the decision?” Locator values need to be comparable across all attributes. If subjective judgments are used across all attributes (for example by using a scale of 0 – 100 with 0 being the “negative” and 100 being the 
“positive” ends of the scale), then, by definition, comparable scaling is used across all attributes. However, if objectively gathered numerical data are used, (for example, if miles/gallon and cost (in dollars) are used as attributes in selecting a vehicle) a scaling approach is needed to assure that the locator values that deem both attributes as “positive” or “negative” are the same \cite{edwards2000multiattribute}. The sixth step is to calculate a utility score for each option. Utility scores are calculated by multiplying the location value of each attribute by its corresponding weight, and then summing across all attributes for each option. The option with the highest utility score is considered the most desirable option and is typically selected as the preferred choice. While this process does not explicitly use equilibrium concepts (game theory), it aims to find a balanced solution where the chosen alternative best satisfies the weighted criteria, akin to reaching an optimal point in a strategic game\cite{von2007theory}. MAUA helps decision-makers to systematically analyze and compare complex decision scenarios involving multiple criteria, providing a structured framework for decision-making that considers the preferences and priorities of stakeholders\cite{keeney1993decisions}. It enables decision-makers to make informed choices that balance trade-offs between different criteria and maximize overall utility or satisfaction. MAUA is computationally efficient when trade-offs and uncertainties are inherent in the decision-making process and where multiple, competing objectives are involved. Utility, or the value associated with a particular decision alternative, is measured as a function of the performance of that alternative on multiple different objectives (see Figure \ref{fig:decision surface}).

MAUA methodology involves the following steps:
\begin{enumerate}
    \item Articulation of decision objective.
    \item Identification of Options: Number of options the user is selecting between 
    \item Identification of Attributes: Key attributes on which the decision will be made
    \item Weight Assignment: Weights are assigned to each attribute based on their importance (user defined)
    \item Utility Locating: Locations are based on each attribute’s performance on each alternative
    \item Aggregation: Scores are aggregated to derive the overall utility of each option
\end{enumerate}

Thus far, the process described has assumed comparability of utilities across attributes using subjective judgments. In many multiattribute utility analyses, various objective (or objective and subjective) utilities are employed. For example, common utilities in vehicle choice would likely include metrics such as miles per gallon (MPG), actual costs, safety ratings, emissions, and comfort and features. MPG, which measures fuel efficiency, is typically evaluated such that higher values are preferred for their indication of better fuel economy. Actual costs encompass the total expenses associated with an option, including purchase price, maintenance, fuel costs, and insurance, with lower costs being more desirable. Safety ratings, derived from agencies like NHTSA or IIHS, assess the protective features and crash worthiness of vehicles, with higher ratings indicating superior safety. Emissions, representing the number of pollutants a vehicle emits, are evaluated, with lower emissions being preferable for environmental and regulatory compliance. Comfort and features are assessed based on the quality of the interior, technological advancements, and overall comfort, where higher values denote a more enjoyable experience. All these attributes are scaled using different, non-comparable metrics.

For effective multi-attribute utility analysis using objective measures, it is crucial to first anchor the reasonably low and reasonably high values for each attribute, establishing a reference range on which to locate the values that link attributes and alternatives. The low value represents the worst or minimum acceptable value, while the high value signifies the best or maximum achievable value. Subsequently, the shape of the utility distribution is specified, with common shapes including linear, concave, convex, and S-shaped distributions. A linear distribution indicates a uniform change in utility with attribute changes, suitable for attributes where each unit change is equally significant. A concave distribution is used where utility increases at a decreasing rate, reflecting diminishing returns. Conversely, a convex distribution indicates utility increases at an increasing rate, fitting attributes where improvements become progressively valuable. The S-shaped distribution combines concave and convex characteristics, apt for attributes with slow initial improvements that accelerate and then decelerate as they approach maximum utility. 

To illustrate, consider evaluating vehicles based on MPG, costs, and safety ratings. Anchoring the low and high values, MPG might range from 15 to 50 with a linear distribution; costs from \$50,000 to \$20,000 with a concave distribution; and safety ratings from 3 to 5 stars with a linear distribution. Utility functions transform these raw values into standardized scores on a 0 to 1 scale. For instance, the MPG utility function could be $U_{MPG} = \frac{MPG - 15}{50 - 15}$, while the actual costs utility function might be $U_{Cost} = \left(1 - \frac{Cost - 20,000}{50,000 - 20,000}\right)^{0.5}$, and the safety ratings utility function $U_{Safety} = \frac{Safety - 3}{5 - 3}$.

Once the low and high anchors are set, and the distribution specified, all objective values can be plotted and their respective standardized locations found on the distribution via interpolation. Individual location scores are then aggregated into a composite utility score for each option using the weighted sum approach as described earlier (we assume weights had been assigned based on the importance of each attribute and add to 1.0). The overall utility score is calculated as $U_{Total} = w_{MPG} \cdot U_{MPG} + w_{Cost} \cdot U_{Cost} + w_{Safety} \cdot U_{Safety}$. This structured approach in multiattribute utility analysis facilitates informed decision-making by balancing different attributes and locations according to the decision maker’s preferences, thereby optimizing the selection process based on multiple criteria.

\section{Methodology}
Multi-attribute utility analysis (MUA) is a decision-making framework that evaluates multiple attributes of a decision option, and is based on the utility function $U$:
\begin{equation}
U = \sum_{i=1}^{n} w_i u_i(x_i)
\end{equation}
Where $n$ is the number of attributes, $w_i$ is the weight of the $i$-th attribute, with $\sum_{i=1}^{n} w_i = 1$. $u_i(x_i)$ is the utility function of the $i$-th attribute, where $x_i$ is the value of the $i$-th attribute. To break it down further, for each attribute $i$:
\begin{equation}
    u_i(x_i) = \texttt{some function of } x_i
\end{equation}
The weights $w_i$ are usually determined through a process such as expert judgment or pairwise comparisons. If we need to incorporate risk or uncertainty, we may use the expected utility approach:
\begin{equation}
E[U] = \sum_{j=1}^{m} p_j U(x_j)
\end{equation}
Where $m$ is the number of possible outcomes, $p_j$ is the probability of the outcome, $j$ and $U(x_j)$ is the utility of the outcome $j$. If the utility function \(u_i(x_i)\) is linear, it simplifies to:

\begin{equation}
U = \sum_{i=1}^{n} w_i x_i
\end{equation}

However, often the utility function is nonlinear, reflecting diminishing or increasing returns on the attribute value. For a more complex case involving interactions between attributes, we might use a multiplicative form:
\begin{equation}
    U = \prod_{i=1}^{n} u_i(x_i)^{w_i}
\end{equation}

To summarize, the general framework in LaTeX for multi-attribute utility analysis can be written as:

\begin{equation}
U = \sum_{i=1}^{n} w_i u_i(x_i)
\end{equation}

Or, under uncertainty:

\begin{equation}
E[U] = \sum_{j=1}^{m} p_j U(x_j)
\end{equation}

To integrate subjective and objective attributes in multiattribute utility analysis (MAUA), it is essential to standardize all attributes to a common scale, facilitating comparison and aggregation. The following method outlines this process:

\begin{enumerate}
    \item Identification of Attributes: Identify all pertinent attributes, both subjective and objective
    \item Determination of Attribute Ranges: Establish the minimum and maximum values for each attribute. Objective attributes often have clearly defined ranges, whereas subjective attributes may require the use of surveys or expert assessments to define their ranges.
    \item Normalization of Attributes: Transform all attributes to a standardized scale, typically ranging from 0 to 100, through normalization. For each attribute \( x_i \), the normalized value \( x_i' \) (where a higher value is desirable) is calculated using:
    \begin{equation}
        x_i' = \frac{x_i - \min(x_i)}{\max(x_i) - \min(x_i)}
    \end{equation} 
    or in the case where a lower value is desirable:
    \begin{equation}
        x_i' = 1 - \frac{x_i - \min(x_i)}{\max(x_i) - \min(x_i)} 
    \end{equation}
    \item Assignment of Weights: The determination of weights can be accomplished via expert judgment, stakeholder consultations, or pairwise comparison techniques.
    \item Calculation of Utility: Compute the utility score by combining the normalized values with their respective weights. The utility score \( U \) for an alternative is calculated as:
    \begin{equation}
         U = \sum_{i=1}^{n} w_i \cdot x_i'
    \end{equation}
    where \( w_i \) represents the weight of the \( i \)-th attribute, and \( x_i' \) is the normalized value of the \( i \)-th attribute.
\end{enumerate}

By employing this standardization process, subjective attributes (such as satisfaction) and objective attributes (such as cost) can be converted to a comparable scale, enabling comprehensive and consistent evaluation in decision-making. This method ensures that all attributes, regardless of their initial measurement units, are assessed uniformly.

A web-based interface, DCZNMaker, that uses the described MAUA methodology is available for use, free of charge. It includes a straightforward and informative user interface (UI), offers robust functionality and features (user can customize the number of options and attributes). DCZNMaker is computationally efficient, and making it available through a public facing website at zero-cost means it can be used readily.

\subsection{Use cases}
Multiattribute utility analysis (MAUA) is a decision-making tool that helps evaluate and compare alternatives based on multiple criteria. Here are three use cases and examples:

Example 1. Healthcare: Choosing a Treatment Plan

A hospital needs to decide on the best treatment plan for patients with a specific condition, considering the attributes (criteria for decision) of: effectiveness (E), side effects (SE), cost (C), patient comfort (PC). The weights are E=(0.4), SE=(0.2), C=(0.2), PC=(0.2). Scoring 3 hypothetical treatment plans: 

\begin{table}[h]
\centering
\begin{tabular}{c|c|c|c|c|c}
\hline
\textbf{Plan} & \textbf{E} & \textbf{SE} & \textbf{C} & \textbf{PC} & \textbf{Utility Score} \\ 
\hline
\hline
Plan A & 80 & 70 & 60 & 90 & 76\\ \hline
Plan B & 85 & 60 & 50 & 85 & 73 \\ \hline
Plan C & 75 & 80 & 70 & 80 & 76\\ \hline
\end{tabular}
\caption{Comparison of treatment plans A, B, and C}
\label{table:plans}
\end{table}

Each plan is assigned a locator score on a scale of 0-100. The utility for each plan is calculated by multiplying the locator score for each criterion by its weight and summing the results. Calculation for plan A: $U_A = 0.4 * 80 + 0.2 * 70 + 0.2 * 60 + 0.2 * 90 = 32 + 14 + 12 + 18 = 76$. Similar calculations are done for Plans B and C, and the plan with the highest score has the highest utility given the importance and locator values. 

Example 2. Personal Finance: Choosing an Investment

An individual is evaluating different investment options based on the attributes: expected return (ER), risk (R), liquidity (L), ethical considerations (EC), with weights ER (0.4), R (0.3), L (0.2), EC (0.1).

\begin{table}[h]
\centering
\begin{tabular}{c|c|c|c|c|c}
\hline
\textbf{Investment Option} & \textbf{ER} & \textbf{R} & \textbf{L} & \textbf{EC} & \textbf{Utility Score}\\ 
\hline
\hline
Option 1 & 85 & 60 & 70 & 90 & 75\\ \hline
Option 2 & 80 & 70 & 80 & 85 & 77.5\\ \hline
Option 3 & 90 & 50 & 60 & 80 & 71\\ \hline
\end{tabular}
\caption{Comparison of Investment Options 1, 2, and 3}
\label{table:investment_options}
\end{table}

Calculation for Option 1 proceeds as  $U_1 = 0.4 \times 85 + 0.3 \times 60 + 0.2 \times 70 + 0.1 \times 90 = 34 + 18 + 14 + 9 = 75$. Similar calculations are done for Options 2 and 3, and the option with the highest utility score is chosen.

In each of these cases, multiattribute utility analysis helps systematically evaluate multiple factors, assign appropriate weights to each, and arrive at a decision that maximizes overall utility.

Example 3 Educational: Evaluating University Programs

A student is choosing the best university program based on four objective attributes: tuition cost (TC), graduation rate (GR), employment rate (ER), and average starting salary (SS). The weights for these criteria are TC (0.3), GR (0.2), ER (0.3), and SS (0.2). Scoring three hypothetical university programs: 

Attributes and Ranges:
\begin{itemize}
    \item Tuition Cost (TC): \$20,000 to \$50,000 per year
    \item Graduation Rate (GR): 50\% to 100\%
    \item Employment Rate (ER): 60\% to 100\%
    \item Average Starting Salary (SS): \$40,000 to \$80,000 per year
\end{itemize}

For a university program with the following values; Program A: TC = \$30,000, GR = 80\%, ER = 90\%, SS = \$60,000, Program B: TC = \$25,000, GR = 70\%, ER = 85\%, SS = \$55,000 and Program C: TC = \$45,000, GR = 90\%, ER = 95\%, SS = \$75,000. To normalize calculations where lower values are desired (TC), we subtract from unity (equation 9). Therefore, for TC with respect to Program A is: \(1 - \frac{30,000 - 20,000}{50,000 - 20,000} = 1 - \frac{10,000}{30,000} = 0.6667\), Program B: \(1 - \frac{25,000 - 20,000}{50,000 - 20,000} = 1 - \frac{5,000}{30,000} = 0.8333\), and Program C: \(1 - \frac{45,000 - 20,000}{50,000 - 20,000} = 1 - \frac{25,000}{30,000} = 0.1667\). To normalize calculations where higher values are desired (GR, ER, SS), we normalize based on equation 8. 

Therefore, the GR for Program A: \(\frac{80 - 50}{100 - 50} = 0.6\), GR for Program B: \(\frac{70 - 50}{100 - 50} = 0.4\), and GR for Program C: \(\frac{90 - 50}{100 - 50} = 0.8\). The ER for Program A: \(\frac{90 - 60}{100 - 60} = 0.75\), ER for Program B: \(\frac{85 - 60}{100 - 60} = 0.625\), and ER for Program C: \(\frac{95 - 60}{100 - 60} = 0.875\). SS for Program A: \(\frac{60,000 - 40,000}{80,000 - 40,000} = 0.5\), SS for Program B: \(\frac{55,000 - 40,000}{80,000 - 40,000} = 0.375\) and SS for Program C: \(\frac{75,000 - 40,000}{80,000 - 40,000} = 0.875\). Using weights: TC (0.3), GR (0.2), ER (0.3), SS (0.2) the utility score is calculated as  $U = (w_{TC} \cdot TC') + (w_{GR} \cdot GR') + (w_{ER} \cdot ER') + (w_{SS} \cdot SS')$. The case of Program A: $U_A = (0.3 \cdot 0.6667) + (0.2 \cdot 0.6) + (0.3 \cdot 0.75) + (0.2 \cdot 0.5) = 0.2 + 0.12 + 0.225 + 0.1 = 0.6667$. Repeating this procedure for Programs B and C, we populate Table \ref{table:universities}. 

\begin{table}[h]
\centering
\begin{tabular}{c|c|c|c|c|c}
\hline
\textbf{Program} & \textbf{Normalized TC} & \textbf{Normalized GR} & \textbf{Normalized ER} & \textbf{Normalized SS} & \textbf{Utility Score} \\ 
\hline
\hline
Program A & 0.6667 & 0.6 & 0.75 & 0.5 & 0.6667 \\ \hline
Program B & 0.8333 & 0.4 & 0.625 & 0.375 & 0.6483 \\ \hline
Program C & 0.1667 & 0.8 & 0.875 & 0.875 & 0.6183 \\ \hline
\end{tabular}
\caption{Comparison of university programs A, B, and C}
\label{table:universities}
\end{table}

Each program's utility score is calculated by multiplying the normalized value for each criterion by its corresponding weight and summing the results. Calculation for Program A:

Similar calculations are done for Programs B and C. The program with the highest utility score is selected as the best option. Based on these calculations, Program A has the highest utility score, making it the best choice for the student according to the weighted criteria. This systematic approach helps the student objectively evaluate and compare the different colleges, considering all important factors.

Example 4. Consumer Electronics: Selecting a Smartphone

A consumer needs to choose the best smartphone based on a mix of objective and subjective attributes. The criteria for decision include cost (C), battery life (BL), camera quality (CQ), and user satisfaction (US). The weights for these criteria are C (0.3), BL (0.3), CQ (0.2), and US (0.2). Scoring three hypothetical smartphone options:

Attributes and Ranges:
\begin{itemize}
    \item Cost (C): \$500 to \$1500
    \item Battery Life (BL): 10 hours to 30 hours
    \item Camera Quality (CQ): 1 to 10 (based on expert reviews)
    \item User Satisfaction (US): 1 to 10 (based on survey data)
\end{itemize}

For three hypothetical smartphone options with the following values, Option A: C = \$1000, BL = 20 hours, CQ = 8, US = 7, Option B:** C = \$700, BL = 25 hours, CQ = 7, US = 8, and Option C: C = \$1300, BL = 15 hours, CQ = 9, US = 6. The normalization calculations are (lower C is better): Option A: \(1 - \frac{1000 - 500}{1500 - 500} = 1 - \frac{500}{1000} = 0.5\), Option B: \(1 - \frac{700 - 500}{1500 - 500} = 1 - \frac{200}{1000} = 0.8\), and Option C: \(1 - \frac{1300 - 500}{1500 - 500} = 1 - \frac{800}{1000} = 0.2\). For attributes BL, CQ, US (where higher is better): BL for Option A: \(\frac{20 - 10}{30 - 10} = 0.5\), BL for Option B: \(\frac{25 - 10}{30 - 10} = 0.75\) and BL for Option C: \(\frac{15 - 10}{30 - 10} = 0.25\). CQ for Option A: \(\frac{8 - 1}{10 - 1} = 0.7778\), CQ for Option B: \(\frac{7 - 1}{10 - 1} = 0.6667\), and CQ for Option C: \(\frac{9 - 1}{10 - 1} = 0.8889\). US for Option A: \(\frac{7 - 1}{10 - 1} = 0.6667\), US for Option B: \(\frac{8 - 1}{10 - 1} = 0.7778\) and US for Option C: \(\frac{6 - 1}{10 - 1} = 0.5556\).

Each option's utility score is calculated by multiplying the normalized value for each criterion by its corresponding weight and summing the results. Calculation for Option A is performed as: 
$U_A = (0.3 \cdot 0.5) + (0.3 \cdot 0.5) + (0.2 \cdot 0.7778) + (0.2 \cdot 0.6667) = 0.15 + 0.15 + 0.1556 + 0.1333 = 0.5889$. Similar calculations are performed for Options B and C, shown in Table \ref{table:smartphones}. The best option with the highest utility score is selected as the best smartphone. 

\begin{table}[h]
\centering
\begin{tabular}{c|c|c|c|c|c}
\hline
\textbf{Option} & \textbf{Normalized C} & \textbf{Normalized BL} & \textbf{Normalized CQ} & \textbf{Normalized US} & \textbf{Utility Score} \\ 
\hline
\hline
Option A & 0.5 & 0.5 & 0.7778 & 0.6667 & 0.5889 \\ \hline
Option B & 0.8 & 0.75 & 0.6667 & 0.7778 & 0.7400 \\ \hline
Option C & 0.2 & 0.25 & 0.8889 & 0.5556 & 0.4667 \\ \hline
\end{tabular}
\caption{Comparison of smartphone options A, B, and C}
\label{table:smartphones}
\end{table}

\begin{figure}
    \centering
    \includegraphics[width=0.8\textwidth]{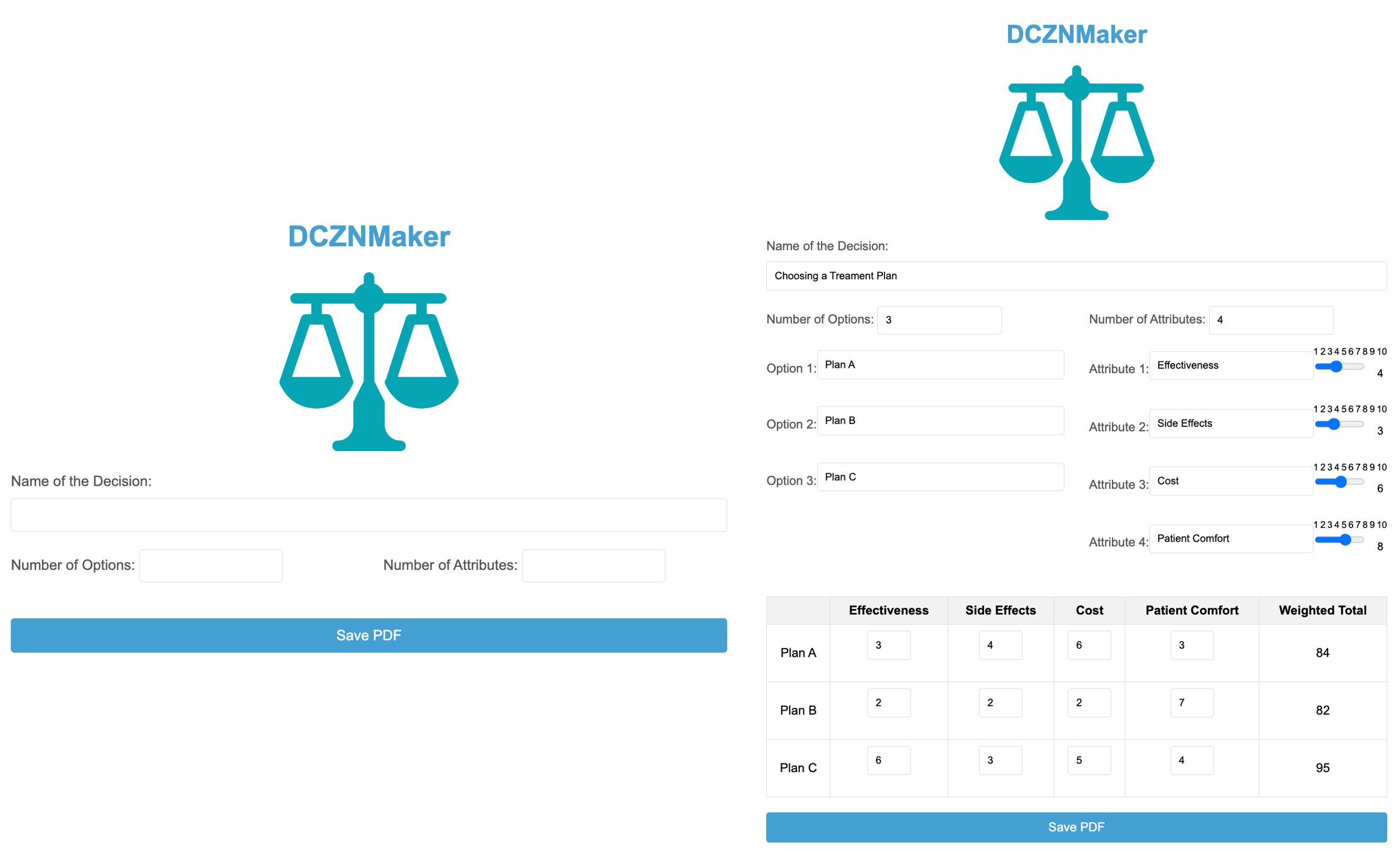}
    \caption{User interface for the application}
    \label{fig: UI}
\end{figure}

The user interface for the web-based implementation of MAUA (DCZNMaker) appears as shown in Figure \ref{fig: UI}.

\section{Discussion}

Multi-Attribute Utility Analysis (MAUA) is a decision-making tool used to evaluate and compare multiple options based on several criteria. As the number of attributes and alternatives increases, the cognitive load on decision-makers becomes substantial. The process of defining utilities, weights, and the relationships between attributes can be overly complex and time-consuming, making it difficult for users to maintain consistency and accuracy, MAUA offers an option to standardize and support these endeavors \cite{reilly2001making}. Its utility spans multiple domains including business, public policy\cite{edwards1977use}, healthcare, engineering and project management, consumerism\cite{van2005predicting}, educational and career planning and finance and investment.

For example, in business, product designs or features are evaluated based on criteria such as cost, market potential, technical feasibility, and customer preferences. This also extends to choosing suppliers or service providers by assessing factors like price, quality, reliability, and service level. In healthcare, DCZNMaker can be used to compare different medical treatments based on effectiveness, side effects, cost, patient preference, and recovery time. It can also aid in resource allocation within a healthcare system by considering factors like patient outcomes,
cost-effectiveness, and accessibility.

In engineering project management, DCZNMaker can assist in selecting engineering projects by evaluating criteria such as technical feasibility, cost, risk, and return on investment. For system design (e.g., transportation networks, IT systems), it balances multiple attributes; performance, reliability, cost, and scalability. In public or environmental policy and planning, DCZNMaker can help assess various urban development projects considering factors such as environmental impact, cost, social benefits, and infrastructure requirements. It is useful for evaluating policies for environmental
protection by weighing criteria like economic impact, sustainability, public health, and regulatory compliance. In the financial sector, DCZNMaker can aid in selecting and evaluating investment portfolios based on risk, return, liquidity, and ensure alignment with investor goals. It can also assist in assessing financial risks by considering various factors such as market volatility, credit risk, and regulatory impact. For consumer decision-making in the purchase of products or services (e.g., cars, homes, appliances), DCZNMaker compares attributes like price, quality, features, and brand reputation. This area extends to travel, where it can help select destinations or packages based on criteria like cost, convenience, activities, and safety. In educational and career planning, DCZNMaker can assist students in choosing colleges by evaluating factors such as academic programs, location, cost, campus facilities, and employment outcomes. It can help individuals make career decisions by considering attributes like job satisfaction, salary, growth opportunities, and work-life balance.
For technology adoption, DCZNMaker can aid in choosing software solutions by evaluating usability, functionality, cost, support, and integration capabilities. It can also help in selecting hardware for an organization by assessing performance, cost, reliability, and vendor support. In all these use cases, MAUA helps decision-makers systematically evaluate and compare alternatives based on a comprehensive set of criteria, leading to more informed and balanced decisions.

Transparency is a strength of MAUA. The process of determining utility functions and weights allow users to visualize the effects of weights and scores creating decision-making transparency, making it easy for others to understand and trust the results. Multi-Attribute Utility Analysis (MAUA) is a powerful tool for decision-making involving multiple criteria, but it has several limitations. Understanding these limitations can help in applying the method appropriately and interpreting its results correctly. One significant limitation is the complexity and data requirements of MAUA. The method is data-intensive, requiring detailed and accurate data on all attributes and their utility functions, which can be time-consuming and costly to collect. Additionally, the complex calculations involved, especially when dealing with many attributes or non-linear utility functions, add to its complexity. Subjectivity is another critical limitation \cite{keeney1993decisions}. Defining utility functions for each attribute involves subjective judgments \cite{store2001integrating}, which can introduce bias. Similarly, determining the weights for different attributes is inherently subjective and may not accurately reflect the true importance of each attribute. MAUA often assumes that attributes are independent of each other, an assumption that may not hold in real-world scenarios where attributes may be interdependent, and their interactions can affect the overall utility \cite{belton2012multiple}. This assumption of independence can lead to inaccurate results. The method also assumes linearity in attribute contributions to overall utility through the additive form of utility functions. However, this linearity assumption may not hold in cases where the relationship between attributes and overall utility is non-linear. MAUA typically assumes that preferences and utilities are static over time, which is a limitation given that in many decision-making contexts, preferences may change due to new information or changing circumstances. This
static nature can affect the relevance of the results. Handling uncertainty and risk is another limitation. MAUA may not adequately address the uncertainty and risk associated with attribute values and their impacts. While some extensions of MAUA incorporate stochastic elements, these add additional complexity. MAUA also assumes a certain level of risk neutrality or risk aversion, which may not accurately reflect the decision maker’s true risk preferences. MAUA can be highly sensitive to input values, such as the utility functions and weights. Small changes in these inputs can lead to significantly different outcomes, necessitating thorough sensitivity analysis to ensure robustness of the results. Finally, ethical and equity considerations are often overlooked in MAUA. MAUA focuses on maximizing utility, which may lead to outcomes that are not socially or ethically desirable. It does not inherently account for ethical or equity considerations, which can be critical in decision-making contexts \cite{greco2016multiple}. Eliciting preferences and translating those preferences into numerical values can be difficult. 

Despite these limitations, MAUA remains a valuable tool for structuring and analyzing complex decision problems involving multiple criteria. Being aware of its limitations allows decision-makers to apply the method more effectively and interpret the results with appropriate caution. DCZNMaker can be access via the URL: \url{https://dcznmaker.org}.

\section{Conclusion}
In this paper, we introduced DCZNMaker, a novel web-based application designed to assist users in making structured decisions through multi-attribute utility analysis (MAUA). By integrating MAUA into a user-friendly platform, DCZNMaker provides a systematic approach for evaluating and comparing multiple decision options based on various criteria and user-defined preferences. DCZNMaker effectively simplifies the decision-making process by breaking down complex decisions into manageable components, allowing users to assign weights to different attributes and
calculate utility scores for each option. The app’s intuitive interface and robust computational capabilities ensure that users can make well-informed decisions quickly and efficiently. It offers the ability to visualize trade-offs and compare alternatives in a structured manner, where dynamic changes in weights or scores facilitate explainable, decision-making.

In conclusion, DCZNMaker represents a significant advancement in decision support tools, offering practical benefits for individuals faced with complex choices. Future developments could include expanding the range of decision contexts, incorporating real-time data integration, and enhancing collaborative decision-making features. As technology continues to evolve, applications like DCZNMaker have the potential to become indispensable tools for decision-makers across various domains.

\section*{Acknowledgments}
This work was inspired by Dr. Theresa Kline.

%Bibliography
\bibliographystyle{unsrt}  
\bibliography{references}

\end{document}